\documentclass[conference]{IEEEtran}

\usepackage{cite, graphicx}
\graphicspath{{fig/}}
\usepackage{hyperref}
\usepackage[cmex10]{amsmath}
\usepackage{amssymb}
\usepackage{siunitx}
\ifCLASSOPTIONcompsoc
\usepackage[caption=false,font=normalsize,labelfont=sf,textfont=sf]{subfig}
\else
\usepackage[caption=false,font=footnotesize]{subfig}
\fi
\usepackage{algorithmic}

\newcommand{\E}[1]{\mathbb{E}\{#1\}}
\newcommand{\e}{\mathrm{e}}
\newcommand{\argmin}{\operatorname{argmin}}
\newtheorem{proposition}{Proposition}

\begin{document}
\title{Energy-Delay Tradeoffs of Virtual Base Stations\\
With a Computational-Resource-Aware\\
Energy Consumption Model}
\author{\IEEEauthorblockN{Tao Zhao, Jian Wu, Sheng Zhou, and Zhisheng Niu}
\IEEEauthorblockA{%
Tsinghua National Laboratory for Information Science and Technology\\
Department of Electronic Engineering, Tsinghua University, Beijing 100084, China\\
Email: \{t-zhao12,wujian09\}@mails.tsinghua.edu.cn,\{sheng.zhou,niuzhs\}%
@tsinghua.edu.cn}}
\maketitle

\begin{abstract}

  The next generation (5G) cellular network faces the challenges of efficiency,
  flexibility, and sustainability to support data traffic in the mobile
  Internet era.  To tackle these challenges, cloud-based cellular architectures
  have been proposed where virtual base stations (VBSs) play a key role.
  VBSs bring further energy savings but also demands a new energy
  consumption model as well as the optimization of computational resources.
  This paper studies the energy-delay tradeoffs of VBSs with delay
  tolerant traffic. We propose a computational-resource-aware
  energy consumption model to capture the total energy consumption of a VBS
  and reflect the dynamic allocation of computational resources including the
  number of CPU cores and the CPU speed.
  Based on the model, we analyze the energy-delay tradeoffs of a VBS considering
  BS sleeping and state switching cost to minimize the weighted sum of
  power consumption and average delay.
  We derive the explicit form of the
  optimal data transmission rate and find the condition under which the energy
  optimal rate exists and is unique. Opportunities to
  reduce the
  average delay and achieve energy savings simultaneously are observed. We further
  propose an efficient
  algorithm to jointly optimize the data rate and the number of CPU cores. Numerical
  results validate our theoretical analyses and under a typical simulation
  setting we find more than 60\% energy savings can be achieved by VBSs
  compared with
  conventional base stations under the EARTH model, which demonstrates the great potential of VBSs in 5G
  cellular systems.

\end{abstract}


\section{Introduction}
\label{sec:intro}

The next generation (5G) cellular network has been attracting research efforts
from both academia and industry.
The requirements and challenges can be summarized as
follows.  First, it is estimated that 5G needs to support 1000 times increase
in traffic capacity~\cite{4gamericas2013meeting}. With limited spectrum and
energy, it is challenging for cellular systems to increase the spectral
efficiency and energy efficiency to cope with the huge traffic demand. Second, 5G
is expected to support massive connections including not only human-to-human
connections but also machine-to-machine connections. Some
of them demand high data rate, while others have loose capacity
requirement but require real time response and high reliability. Hence, the
cellular network must be flexible enough to adapt to different connections with
different characteristics. Besides, with the influence of innovative
applications from IT companies, the average revenue per user of
network operators tends to
increase slowly or even decrease in some cases, while the expenditures increase
rapidly~\cite{cmri2013cran}. Such trend imposes a great challenge to the
sustainability of the cellular network.

Facing these challenges, conventional cellular architectures can hardly support
the 5G systems for the following reasons. First, in conventional cellular
architectures, resources are commonly provisioned according to peak traffic
requirement. While this approach ensures quality of service (QoS), it inevitably wastes a
lot of resources in realistic networks where data traffic is highly dynamic.
Second, conventional base stations (BSs) manage each cell in a distributed
manner. The lack of BS cooperation results in inflexibility, and makes it
difficult to utilize coordinated multi-point communications (CoMP) and
coordinated BS sleeping to increase the spectral efficiency and energy efficiency.
Moreover, conventional BS is a complicated system where hardware and software
are tightly coupled, so operators cannot easily upgrade it, nor deploy value
added services quickly. To sum up, it is difficult for conventional cellular
architectures to tackle the challenges faced by 5G including efficiency,
flexibility, and sustainability.  Therefore it is crucial to renovate
cellular network architectures to meet the requirements of 5G systems.

One of the promising architecture evolution trends is integrating cloud computing technology into
cellular networks. Wireless Network Cloud (WNC)~\cite{lin2010wireless} proposed
by IBM researchers and CRAN~\cite{cmri2013cran} proposed by China Mobile share
the same idea of moving base band units (BBUs) of BSs to
a centralized cloud computing platform, and only leaving remote radio heads
(RRHs) in the front end. With the help of open IT platforms,
cellular systems can be more flexible and sustainable. The deployment of CRAN
also demonstrates its capability to reduce the cost and improve radio access
performance.  Based on the existing research, we have proposed
CONCERT~\cite{liu2014concert}, which stands for CONvergence of Cloud and
cEllulaR sysTems. Its main features are heterogeneous physical resources,
logically centralized resource virtualization, and software defined services.
They altogether allow CONCERT to support
5G cellular systems and provide innovative services.

In cloud-based cellular network architectures, virtualization technology
is vital to make BSs software defined, that is to make them virtual base stations (VBSs). The
main advantage compared with conventional BSs is that computational resources
such as the number of CPU cores and the CPU speed
of VBSs are pooled and can be dynamically allocated to each VBS to adapt to
the dynamics of the traffic demand over time and space, which brings
further energy savings. However, computational resource dynamics are
not captured in the existing BS energy consumption
models~\cite{auer2011howmuch,gupta2012energy}. Therefore we propose a new
model to assist the research, which, to our best knowledge, is the first
computational-resource-aware energy consumption model for VBSs.

Energy-delay tradeoffs of BSs in wireless systems have been
studied in many literatures. It was pointed out that when taking practical
concerns into account, the energy-delay tradeoff deviates from the simple
monotonic curve~\cite{chen2011fundamental}.  Our previous work analyzed the
energy-delay tradeoffs of conventional BSs with the EARTH energy consumption
model~\cite{wu2012traffic,wu2013traffic}.  In this paper we analyze the
energy-delay tradeoff relationship of VBSs with a computational-resource-aware
model considering BS sleeping and state switching cost.
Moreover, we investigate the impact of computational resources on the
relationship, and compare
the energy saving performance of VBSs in cloud-based architectures with
EARTH BSs in conventional cellular networks to show the
energy saving gain of VBSs.

The main contributions of the paper are as follows.
\begin{enumerate}
  \item We propose a computational-resource-aware energy consumption model for
    VBSs which can capture the total energy consumption and
    reflect the dynamic allocation of computational resources.

  \item We derive the explicit form of the optimal data transmission rate
    which minimizes the weighted sum of power consumption and average delay,
    and find the condition under which the energy optimal rate exists. This
    property
    indicates the opportunity to reduce the average delay and save energy
    simultaneously.

  \item We investigate the impact of computational resources on energy-delay
    tradeoffs of VBSs and propose an efficient algorithm to optimize the data
    rate and the number of CPU cores jointly.

\end{enumerate}

The rest of this paper is organized as follows. We first present our
computational-resource-aware energy consumption model in
Section~\ref{sec:model}. Then we
describe the system model in Section~\ref{sec:sysmodel}.  Theoretical analyses
of energy-delay tradeoffs are given in Section~\ref{sec:edt}.  We show our
numerical results in Section~\ref{sec:num} and then conclude the paper in
Section~\ref{sec:con}.

\section{Energy Consumption Model}
\label{sec:model}
\subsection{EARTH Model}
The EARTH energy consumption model of BSs~\cite{auer2011howmuch}
has been widely adopted in
the literature to analyze the energy efficiency of cellular systems.
It has the form as:
\begin{equation}
  P_\text{in} = \begin{cases}
    N_{\text{TRX}} P_0 + \Delta_\text{p} P_\text{out}, & 0 < P_\text{out} \le P_\text{max} \\
    N_{\text{TRX}} P_\text{sleep}, & P_\text{out} = 0
  \end{cases}
  \label{eqn:earth}
\end{equation}
where $P_\text{in}$ is the total power supply of the BS,
and $P_\text{out}$ is the output power per antenna measured at the input of the
antenna element.
$N_{\text{TRX}}$ is the number of antennas at the BS,
$P_0$ is the power consumption at the minimum non-zero load,
$\Delta_\text{p}$ is the slope of load varying power consumption,
and $P_\text{sleep}$ is the energy consumption in sleep mode.

This model cannot be directly used with VBSs for two reasons.
One is that multiple BBUs reside in one cloud infrastructure,
so the energy consumption of BBU per BS should be reduced.  The
other is that by virtualization, the base band computational resources can be
dynamically allocated, and the BBU application can be run only when
necessary. However, the EARTH model cannot reflect the variations of
computational resources. As a result, a new
model for VBSs under cloud-based cellular architectures
is required.

\subsection{Computation Resource Aware Model}

Based on the analysis of the existing energy consumption model, we propose a
computational-resource-aware energy consumption model for VBSs.
Following the component based methodology of the EARTH model, we
calculate the power consumption of the BBU and the RRH in the VBS separately,
and take the summation as the total power consumption:
\begin{equation}
  P = P_\text{R} + P_\text{B}
  \label{eqn:total-power}
\end{equation}
where $P_\text{R}$ and $P_\text{B}$ are the power consumption of the RRH and the BBU respectively.

Regarding the RRH part, we leverage the intermediate
result from the EARTH model~\cite{auer2011howmuch}:
\begin{equation}
  P_\text{R} = \frac{P_\text{out}}{\eta}+P_\text{RF}
  \label{eqn:rrh-power}
\end{equation}
where $\eta$ denotes the power amplifier (PA) efficiency,
and $P_\text{RF}$ denotes the
power consumption of the radio frequency (RF) circuits.

As for the BBU part, we calculate the energy consumption as follows:
\begin{equation}
  P_\text{B} = N_\text{c}(P_\text{Bm}+ \Delta_{P_\text{B}} \rho_\text{c} s^\beta)
  \label{eqn:pbs}
\end{equation}
where
\begin{equation}
  \Delta_{P_\text{B}} = (P_\text{BM}-P_\text{Bm}) / s_0^\beta ,
  \label{eqn:Delta}
\end{equation}
$N_\text{c}$ denotes the number of \emph{active} CPU cores,
$P_\text{Bm}$ and $P_\text{BM}$ are the minimum and maximum power consumption of
each core,
$\rho_\text{c}$ denotes the CPU load by the BBU process for $N_\text{c}$
cores which is usually expressed in percentage,
$s$ is the CPU speed,
$s_0$ is the reference CPU speed,
and $\beta$ is the exponential coefficient of CPU speed.
In this model
the power consumption of the BBU is linear with the number of CPU cores as well as
the CPU load~\cite{blackburn2008five,vasan2010worth}.
With speed-scaling, the dynamic part of the power
consumption is polynomial with the CPU speed besides the CPU
load~\cite{son2012speedbalance}.

Furthermore, we capture the relationship between the utilized computational
resources and the software tasks which compute samples for wireless
communications.
The CPU load can be expressed by the following equation:
\begin{equation}
  \rho_\text{c} = \frac{f(r)}{N_\text{c} s} = \frac{c_0+\kappa r}{N_\text{c} s}
  \label{eqn:rho_c}
\end{equation}
where $f(r)$ is the actual instructions per unit
time and $N_\text{c} s$ represents the maximum instructions available per unit time.
We assume $f(r)$ is linear with the data transmission rate $r$, where
$c_0$ and $\kappa$ are relevant coefficients.
The assumption is based on the profiling result
in CloudIQ~\cite{bhaumik2012cloudiq} which shows
a linear relationship between
the processing time of an LTE subframe and the modulation and coding scheme (MCS)
used as well as the physical resource blocks (PRB) available.

Substitute (\ref{eqn:rho_c}) into (\ref{eqn:pbs}), we get
\begin{equation}
  P_\text{B} = N_\text{c} P_\text{Bm} + \Delta_{P_\text{B}}c_0s^{\beta - 1} +
  \Delta_{P_\text{B}} \kappa r s^{\beta - 1}
  \label{eqn:bbu-power-s}
\end{equation}
which means the computational power consumption is linear with the data rate.

In summary, the power consumption of a VBS is:
\begin{equation}
  P = \begin{cases}
    P_\text{B} +  P_\text{R}, & 0 < P_\text{out} \le P_\text{max}\\
    P_\text{sleep}, & P_\text{out} = 0
  \end{cases}
  \label{eqn:model}
\end{equation}

\section{System Model}
\label{sec:sysmodel}

We consider one VBS on a server with $N_\text{c}$ active CPU cores with speed $s$.
We model the system as an M/G/1 Processor Sharing (PS) queue.
Traffic flows arrive at the BS with average rate $\lambda$, and each flow has
an average file size $L$.
The data transmission rate is $r$ \si{bps} when the queue has customers;
otherwise it is zero.
So the traffic load of the queue is $\rho = \lambda L / r $.
According to queueing theory, the average queue length is:
\begin{equation}
  \E{n} = \frac{\rho}{1-\rho} = \frac{\lambda L}{r - \lambda L}
  \label{eqn:En}
\end{equation}
By applying Little's Law, we know the average delay is $\E{D} = \E{n} / \lambda$.

In our model the VBS will go to sleep when there is no customer in the queue (``off''
state), and be back to work when a new customer arrives (``on'' state).
In an on-off cycle, we let $T_\text{a}$ denote the time
duration in the busy period, $T_\text{s}$ the time in the consecutive idle period,
and $T_\text{c} = T_\text{a} + T_\text{s}$ the total time.
We assume a switching cost $E_\text{sw}$ is incurred during each on-off state
switching, so the average power consumption in a cycle is as follows:
\begin{equation}
  \E{P} = p_\text{active} (P_\text{B} + P_\text{R}) + p_\text{sleep} P_
  \text{sleep} + \frac{2 E_\text{sw}}{\E{T_\text{c}}}
  \label{eqn:Ep}
\end{equation}
where $p_\text{active}=\rho$ and $p_\text{sleep}=1-\rho$ are the fraction of time of busy
and idle period during one cycle respectively.

As for the cell coverage, we adopt the standard 3GPP propagation model.
The radius of cell is $R$. We consider large scale path loss while ignoring
shadowing loss. The downlink signal-to-interference-plus-noise ratio (SINR)
is given by:
\begin{equation}
  \text{SINR}(d) = g P_{\text{out}} =
  \frac{P_{\text{out}}}{L(d) F N_0 W}
  \label{eqn:sinr-g}
\end{equation}
where $L(d)$ is the path loss, $F$ is the noise factor of the user equipment
(UE), $N_0$ is the
noise spectral density, and $W$ is the system bandwidth.
For explicit analysis, we assume all users are located at the cell
edge, so the overall channel gain $g$ is the same for each user.
Therefore the sum data rate at the BS can be expressed by the following
equation:
\begin{equation}
  r = W \log_2 (1+ g P_{\text{out}})
  \label{eqn:rate-g}
\end{equation}

The optimization problem is as follows:
\begin{align}
  \min_{r, N_\text{c}}\quad & z = \mathbb{E}\{P\} + \alpha \E{n}
\end{align}
We want to minimize the system cost $z$, which is a weighted sum of average power
consumption and average queue length. $\alpha$ is the weighting factor.
The decision variables are the data rate $r$ and the number of CPU cores
$N_\text{c}$.

\section{Energy-Delay Tradeoffs}
\label{sec:edt}

To get the optimal solution to the above problem, we first let
$ \frac{\partial z}{\partial r} = 0 $, and find that
the optimal rate $r^*$ satisfies:
\begin{equation}
  \Omega\left( \frac{\alpha g \eta}{\e} \left(\frac{r^*}{r^*
  - \lambda L}\right)^2 + \frac{g \eta P_\text{s} - 1}{\e}
  \right) = \frac{r^* \ln 2}{W} - 1
  \label{eqn:opt}
\end{equation}
where
\begin{align}
  P_\text{s} &= P_\text{o}-P_\text{sleep}- 2\lambda E_\text{sw}, \\
  P_\text{o} &= N_\text{c} P_\text{Bm} + \Delta_{P_\text{B}} c_0 s^{\beta - 1} + P_{\text{RF}} ,
  \label{eqn:Ps}
\end{align}
and $\Omega(\cdot)$ is the principal branch of Lambert W function.
As $r$ increases,
the left side of Eqn.~(\ref{eqn:opt}) decreases and the right side increases.
Therefore the optimal rate $r^*$ is unique.

We have the following proposition on the energy-delay tradeoff relationship
of VBSs with varying data rate $r$.
\begin{proposition}
  For the relationship between average power consumption and average
  delay, we have:
  \begin{enumerate}
    \item There exists the unique energy optimal rate $r_\text{e}^*$ when the
  following condition is satisfied:
  \begin{align}
    \lambda & < \frac{P_\text{o} - P_\text{sleep}}{2 E_\text{sw}}, \\
    L & < \frac{W}{\lambda \ln 2} \left[ \Omega\left(\frac{g \eta
    P_\text{s}- 1}{\e}
    \right)+ 1 \right]
    \label{eqn:condition}
  \end{align}
  \label{eqn:proposition1}
  The corresponding energy optimal rate is given by:
\begin{equation}
  r_\text{e}^* = \frac{W}{\ln 2} \left[ \Omega\left(\frac{g \eta P_\text{s} - 1}{\e}
  \right)+ 1 \right] .
  \label{eqn:e-opt}
\end{equation}
\item  When the above condition is not satisfied,
  the average power consumption is
  monotonically decreasing with the average delay.
\item In both cases, when the average delay goes to infinity, the average
  power consumption approaches $P_\text{o} + \kappa \Delta_{P_\text{B}} s^{\beta
  - 1} \lambda L + \frac{2^{\frac{\lambda L}{W}} - 1}{g \eta}$ .
  \end{enumerate}
\end{proposition}

The proof is omitted for brevity.

\emph{Remark:}
When the energy optimal point exists, average delay can be traded off
for energy savings only when $r > r_\text{e}^*$, and when $r < r_\text{e}^*$
we can reduce the average delay and save energy simultaneously.
Interestingly, the proposition has the same mathematical structure as that in
our previous work~\cite{wu2013traffic}, and for VBSs
the energy optimal rate is not affected by the part of computational power
consumption that is linear with data rate. The reason is that the effect of
that part of
computational power consumption is neutralized by time fraction factor
influenced by traffic load which is inversely proportional to the data rate.

Further we investigate the impact of computational resources, in particular
the impact of the number of CPU cores.
On one hand, if we fix the number of CPU cores, it sets the maximum supportable
rate:
\begin{equation}
  r_\text{M}(N_\text{c}) = \frac{N_\text{c} s - c_0}{\kappa}
  \label{eqn:rM}
\end{equation}
On the other hand, we have
$ \frac{\partial z}{\partial N_\text{c}} = P_\text{Bm} > 0 $,
$ \frac{\partial r^*}{\partial N_\text{c}} > 0 $,
which means increasing the number of CPU cores
will increase the system cost and the optimal rate.

Based on the above analyses, we propose an efficient algorithm in
Fig.~\ref{fig:alg-pair} to find the optimal data rate and
number of CPU cores $(r^*, N_\text{c}^*)$ jointly.
In the algorithm, we search through the zone of average delay decreasingly.
At first only one CPU core is considered.
When the current number of CPU core cannot support the local optimal rate under
that number, we mark the maximum supportable rate as one candidate of
global optimal solution, and consider one more CPU core.
When we find the number of CPU cores under which the local optimal rate
can be achieved,
we can mark the local optimal rate as the final candidate and exit the search
since the total cost always increases afterwards.
At last the global optimal rate and number of CPU cores
can be obtained by comparing all the candidates.
In this way it is unnecessary to exhaustively search all the possible numbers of CPU cores and make comparisons, which makes the algorithm efficient.

\begin{figure}[!t]
  \centering
  \begin{algorithmic}[1]
    \STATE Set $N_\text{cM}$, $N_\text{c} \leftarrow 1$, $S \leftarrow \Phi$
    \WHILE{$N_\text{c} \le N_\text{cM}$}
    \STATE $\hat{r}(N_\text{c}) \leftarrow \argmin_{r} z(r, N_\text{c})$
    \IF{$\hat{r}(N_\text{c}) \le r_\text{M}(N_\text{c})$}
    \STATE $S \leftarrow S \cup \{(\hat{r}(N_\text{c}), N_\text{c})\}$
    \STATE Break out of the loop
    \ELSE
    \STATE $S \leftarrow S \cup \{(r_\text{M}(N_\text{c}), N_\text{c})\}$
    \STATE $N_\text{c} \leftarrow N_\text{c} + 1$
    \ENDIF
    \ENDWHILE
    \RETURN $(r^*, N_\text{c}^*) = \argmin_{(r, N_\text{c}) \in S} z(r, N_\text{c})$
  \end{algorithmic}
  \caption{The algorithm to find the optimum $(r^*, N_\text{c}^*)$.
  $N_\text{cM}$ is the maximum number of CPU cores in practical systems.
  $S$ is the set to store the candidates of optimal points.
  $\Phi$ is the empty set.}
  \label{fig:alg-pair}
\end{figure}

\section{Numerical Results}
\label{sec:num}

In this section we present the
numerical results to show the energy-delay tradeoffs of VBSs.
The simulation parameters of VBSs are listed in Table~\ref{tab:edt}.
Among them the base band parameters are based on commodity servers, and
the cellular parameters are from LTE R11 standard.
As for the conventional BS under the EARTH model, we set
$P_0 = \SI{84}{W}, \Delta_\text{p} = 2.8, P_\text{sleep} = \SI{56}{W}$,
and $N_\text{TRX} = 1$.

\begin{figure}[!t]
  \centering
  \includegraphics[width=.33\textwidth]{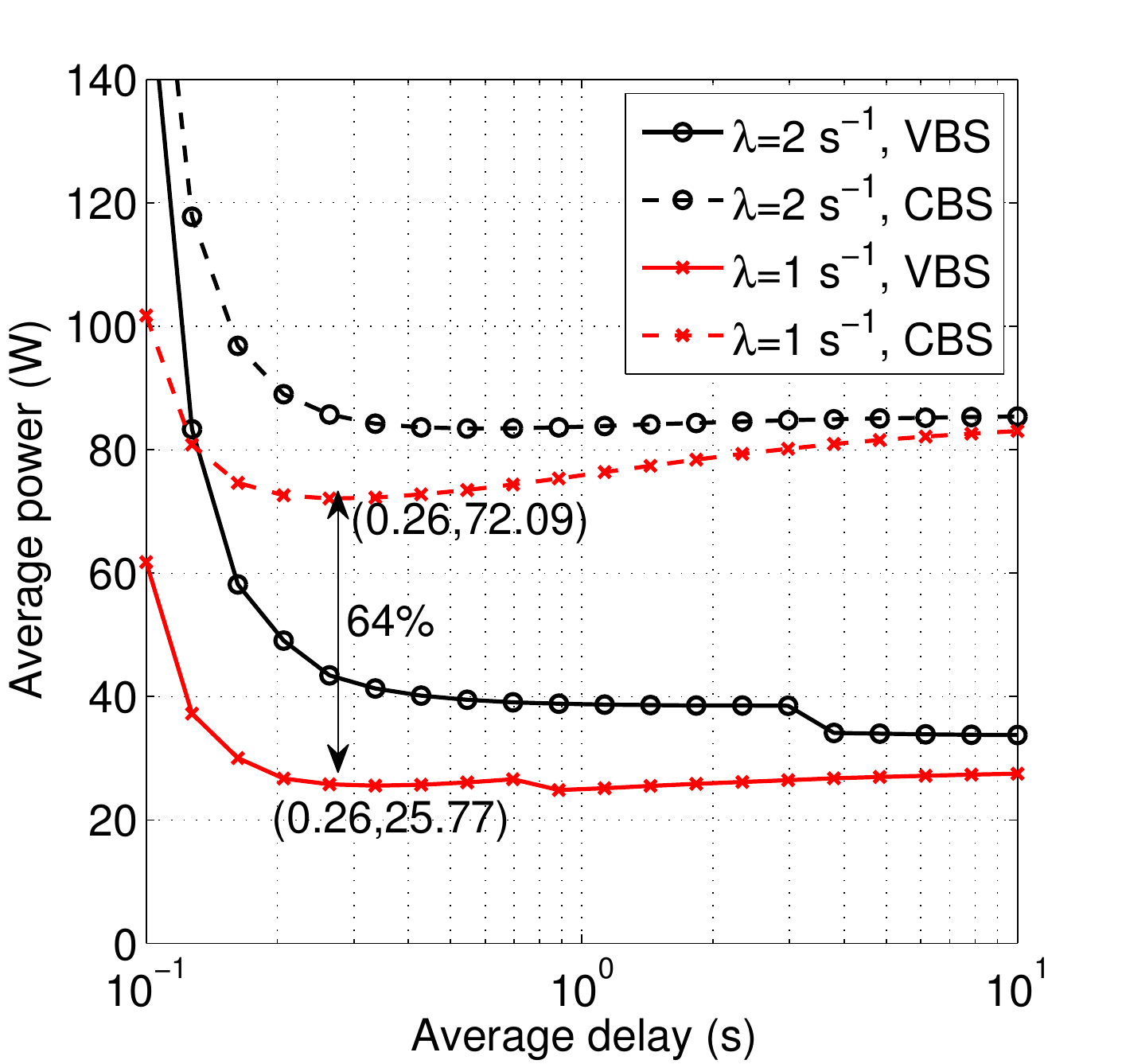}
  \caption{Power consumption comparison between the VBS and the conventional BS
  (CBS).  $L = \SI{2}{MB}$.}
  \label{fig:edt-earth}
\end{figure}

Fig.~\ref{fig:edt-earth} shows the energy saving performance of VBSs
compared with BSs under the EARTH model. We can find more than 60\% energy
consumed by conventional BSs can be saved when using VBSs. For example
when $\lambda = \SI{1}{s^{-1}}$,
about 64\% savings can be achieved with the same average delay $\E{D}=\SI{0.26}{s}$
that optimizes the power consumption of conventional BSs.
The savings come from traffic aware computational power consumption.
The BBU power consumption scales with the actual data rate, rather than stays
static in conventional BSs.

\begin{figure}[!t]
  \centering
  \includegraphics[width=.33\textwidth]{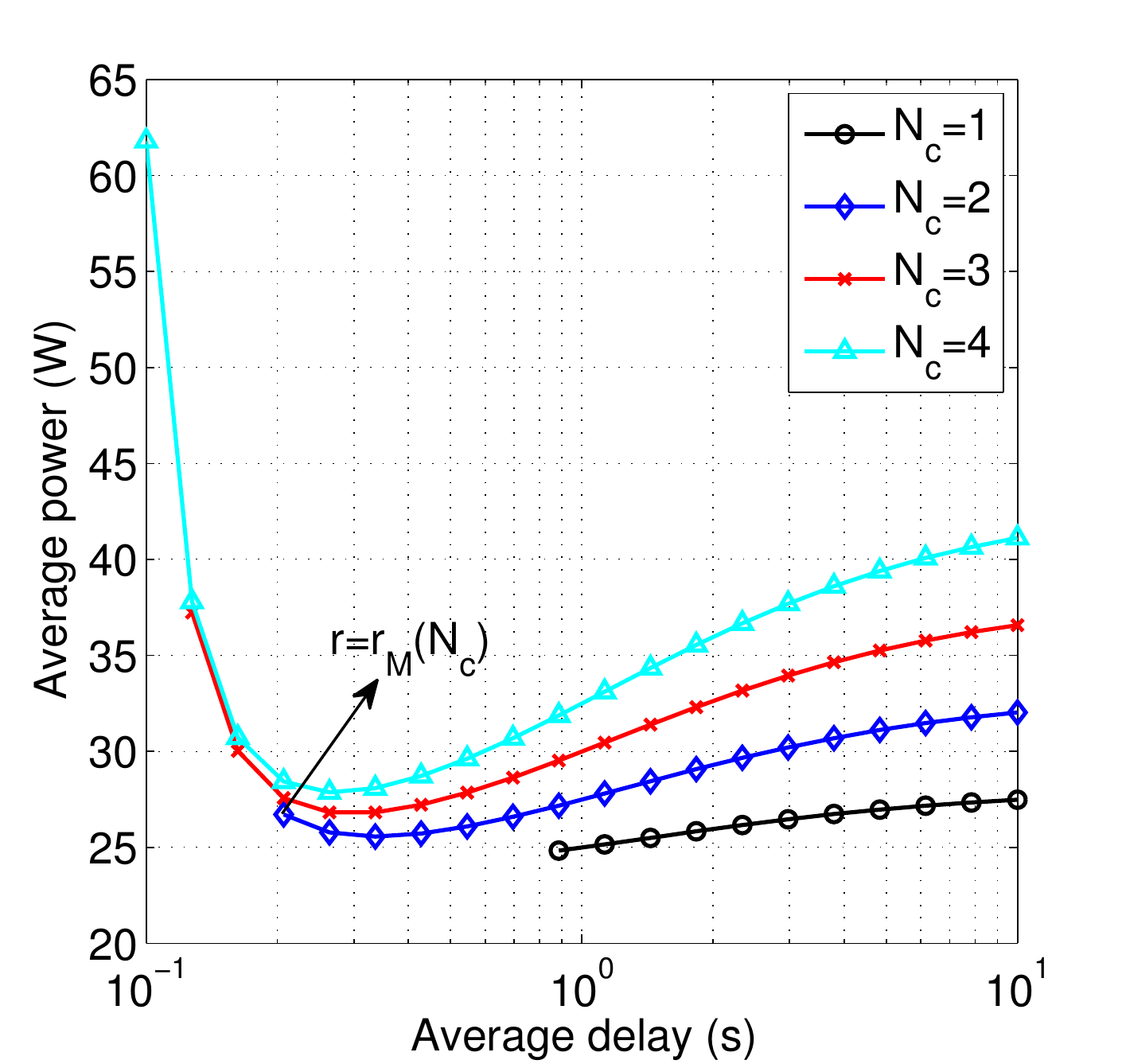}
  \caption[Energy-delay tradeoff with different numbers of CPU cores.]%
  {Energy-delay tradeoff with different numbers of CPU cores $N_\text{c}$.\\
  $\lambda = \SI{1}{s^{-1}}, L = \SI{2}{MB}$.}
  \label{fig:edt-Nc}
\end{figure}

Fig.~\ref{fig:edt-Nc} shows the relationship between average power consumption
and average delay given different numbers of CPU cores.
For example when $N_\text{c} = 4$,
there exists the unique optimal point to minimize the energy consumption.
To the right of the optimal point, there is the
opportunity to reduce the average delay and achieve energy savings
simultaneously.
In addition, the impact of the number of CPU cores on energy-delay tradeoffs is
presented in the figure.
The left end points of the curves for smaller numbers of CPU cores mark the maximum
supportable data rate.
Given the average delay, increasing the number of
CPU cores will increase the average power consumption as well as the energy
optimal rate.

\begin{figure}[!t]
  \centering
  \includegraphics[width=.33\textwidth]{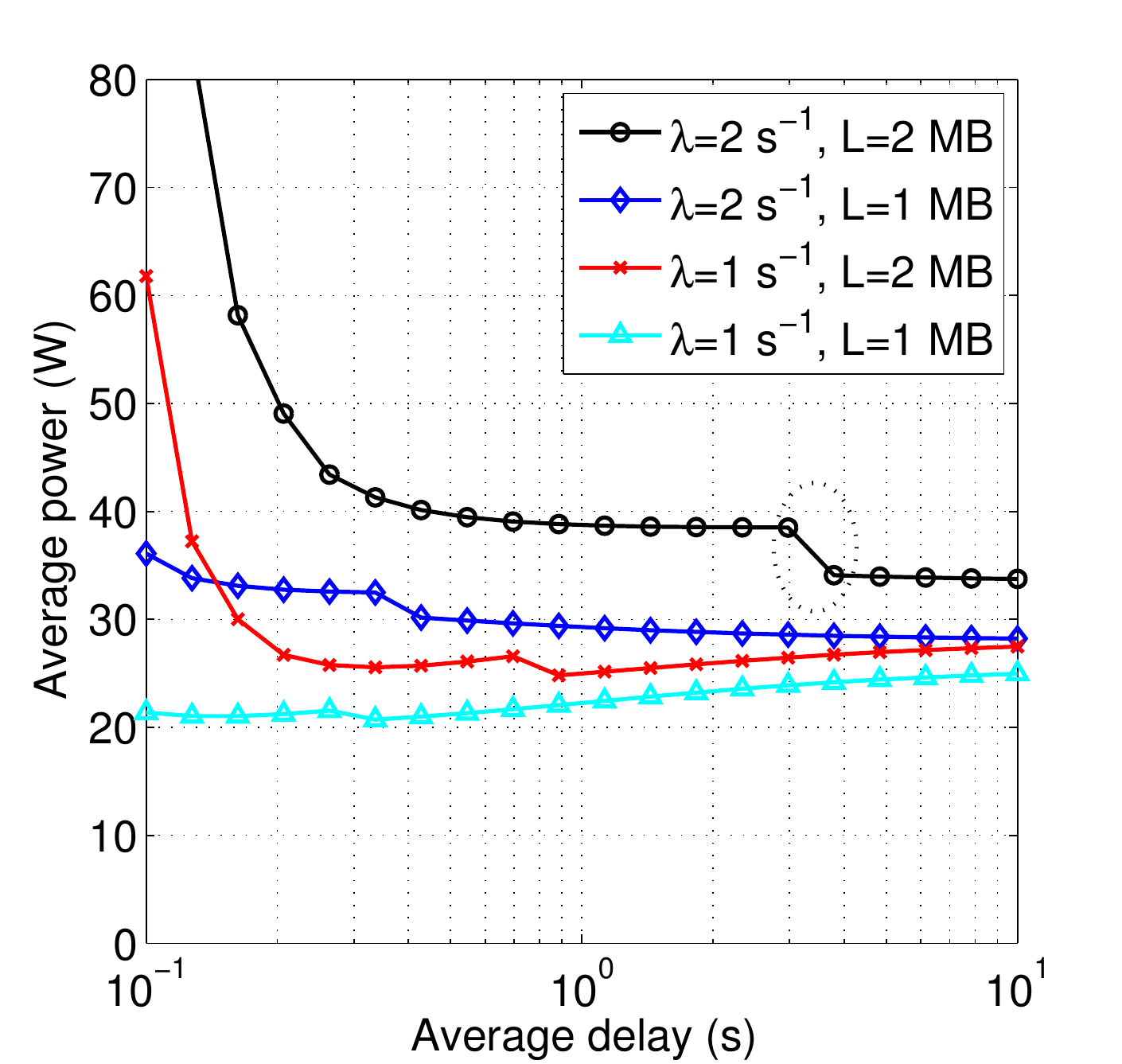}
  \caption{Energy-delay tradeoff with different arrival rates.}
  \label{fig:edt-traffic}
\end{figure}

When both the data rate and the number of CPU cores are adjusted, the
energy-delay tradeoff relationship
between the average power consumption and the average delay is shown with
different traffic load in Fig.~\ref{fig:edt-traffic}. Note
each curve is divided into several zones due to the impact of the number
of CPU cores. The algorithm to find the optimum
$(r^*, N_\text{c}^*)$ can be illustrated by the figure.
Take the red curve with cross markers
as an example. We need to compare the
rightmost turning point with $\E{D}=\SI{0.89}{s}$ and the local optimal point
with $\E{D} = \SI{0.34}{s}$
to determine the global optimal solution.
Besides, Fig.~\ref{fig:edt-traffic} depicts the impact of traffic
arrival on the energy-delay tradeoff.
Either larger arrival rate or larger average file size will increase
the average power consumption given average delay. The power consumption when
average delay approaches infinity is monotonically increasing with $\lambda L$.
Specially the two curves in the middle
with the same $\lambda L$ approach the same asymptotic value.

\begin{table}[!t]
  \renewcommand{\arraystretch}{1.3}
  \caption{Simulation parameters}
  \label{tab:edt}
  \centering
  \begin{tabular}{lc}
    \hline
    CPU speed ($s$) & \SI{2}{GHz} \\
    \hline
    Reference CPU speed ($s_0$) & \SI{2}{GHz} \\
    \hline
    Maximum power per CPU core ($P_\text{BM}$) & \SI{20}{W} \\
    \hline
    Minimum power per CPU core ($P_\text{Bm}$) & \SI{5}{W} \\
    \hline
    Exponential coefficient of CPU speed ($\beta$) & 2 \\
    \hline
    Constant coefficient of instruction speed ($c_0$) & $7\times10^8$ \\
    \hline
    Rate varying coefficient of instruction speed ($\kappa$) & 35 \\
    \hline
    Carrier frequency ($f$) & \SI{2}{GHz} \\
    \hline
    Cell radius ($R$) & \SI{0.5}{km} \\
    \hline
    UE noise figure ($F$ in \si{dB}) & \SI{9}{dB} \\
    \hline
    Noise spectral density ($N_0$) & \SI{-174}{dBm/Hz} \\
    \hline
    System bandwidth ($W$) & \SI{20}{MHz} \\
    \hline
    RF circuit power ($P_\text{RF}$) & \SI{12.9}{W} \\
    \hline
    PA efficiency ($\eta$) & 31.1\% \\
    \hline
    VBS sleeping power ($P_\text{sleep}$) & \SI{6.45}{W} \\
    \hline
    Switch cost ($E_\text{sw}$) & \SI{5}{J} \\
    \hline
  \end{tabular}
\end{table}

\section{Conclusion}
\label{sec:con}

In this paper we propose a computational-resource-aware energy consumption
model for VBSs in cloud-based cellular network architectures, and
investigate the energy-delay tradeoffs of a VBS considering BS sleeping.
We give the explicit form of the optimal data rate, and find the property which
can depict the opportunity
to achieve energy savings and reduce the average delay simultaneously.  We
further investigate the impact of computational resources and propose an
efficient algorithm to jointly optimize the data rate and
the number of CPU cores. Numerical results validate our theoretical analyses and reveal
that more than 60\% energy savings can be brought by VBSs compared with
conventional BSs under the EARTH model.  Future work will consider
the multi-BS scenario.

\section*{Acknowledgment}

The authors would like to thank Dr.~Shugong Xu and Dr.~Shan Zhang for
helpful discussions.
This work is sponsored in part by the National Basic Research Program of China
(973 Program: 2012CB316001), the National Science Foundation of China (NSFC)
under grant No. 61201191, the Creative Research Groups of NSFC under grant No.
61321061, and Intel Corporation.

\end{document}